\documentclass[10pt,superscriptaddress,twocolumn,showpacs,prb,aps]{revtex4-1}

\usepackage{graphicx}
\usepackage{dcolumn}
\usepackage{bm}
\usepackage{xspace}
\usepackage{multirow}
\usepackage{natbib}
\usepackage{color}
\usepackage{makecell}

\newcommand{\RUC}{Department of Physics, Renmin University of China, Beijing 100872, China}
\newcommand{\BESSY}{Institute for Solid State Research, IFW Dresden, Dresden 01171, Germany}
\newcommand{\IOP}{Beijing National Laboratory for Condensed Matter Physics, and Institute of Physics, Chinese Academy of Sciences, Beijing 100190, China}
\newcommand{\CenterQM}{Collaborative Innovation Center of Quantum Matter, Beijing, China}
\newcommand{\BJLab}{Beijing Key Laboratory of Opto-electronic Functional Materials $\textsl{\&}$ Micro-nano Devices, Renmin University of China, Beijing, China}
\newcommand{\Rice}{Department of Physics and Astronomy, Rice University, Houston, Texas 77005, USA}

\newcommand{\PHT}{Pd$_x$HoTe$_{3}$}
\newcommand{\HT}{HoTe$_{3}$}
\newcommand{\RT}{$R$Te$_{3}$}
\newcommand{\EF}{$E_F$}
\newcommand{\cdw}{0.01}
\newcommand{\cdwsc}{0.02}
\newcommand{\SC}{0.04}

\begin{document}

\title{Interplay between multiple charge-density waves and the relationship with superconductivity in {\PHT}}

\author{Rui Lou}
\affiliation{\RUC}
\affiliation{\BJLab}

\author{Yipeng Cai}
\thanks{Present address: Department of Physics, McMaster University, Hamilton, Ontario, L8S 4M1, Canada.}
\affiliation{\RUC}
\affiliation{\BJLab}

\author{Zhonghao Liu}
\thanks{Present address: State Key Laboratory of Functional Materials for Informatic, SIMIT, Chinese Academy of Sciences, Shanghai 200050, China.}
\affiliation{\BESSY}

\author{Tian Qian}
\affiliation{\IOP}

\author{Lingxiao Zhao}
\affiliation{\IOP}

\author{Yu Li}
\affiliation{\Rice}

\author{Kai Liu}
\affiliation{\RUC}
\affiliation{\BJLab}

\author{Zhiqing Han}
\author{Dandan Zhang}
\affiliation{\RUC}
\affiliation{\BJLab}

\author{Junbao He}
\thanks{Present address: Physics and Electronic Engineering College, Nanyang Normal University, Nanyang 473061, China.}
\affiliation{\IOP}

\author{Genfu Chen}
\author{Hong Ding}
\affiliation{\IOP}
\affiliation{\CenterQM}

\author{Shancai Wang}
\email{scw@ruc.edu.cn}
\affiliation{\RUC}
\affiliation{\BJLab}

\begin{abstract}
  {\HT}, a member of the rare-earth tritelluride ({\RT}) family, and its Pd-intercalated compounds, {\PHT}, where superconductivity
  (SC) sets in as the charge-density wave (CDW) transition is suppressed by the intercalation of a small amount of Pd, are investigated
  using angle-resolved photoemission spectroscopy (ARPES) and electrical resistivity. Two incommensurate CDWs with perpendicular nesting
  vectors are observed in {\HT} at low temperatures. With a slight Pd intercalation ($x$ = {\cdw}), the large CDW gap decreases and the
  small one increases. The momentum dependence of the gaps along the inner Fermi surface (FS) evolves from orthorhombicity to near
  tetragonality, manifesting the competition between two CDW orders. At $x$ = {\cdwsc}, both CDW gaps decreases with the emergence of SC.
  Further increasing the content of Pd for $x$ = {\SC} will completely suppress the CDW instabilities and give rise to the maximal SC order.
  The evolution of the electronic structures and electron-phonon couplings (EPCs) of the multiple CDWs upon Pd intercalation are carefully
  scrutinized. We discuss the interplay between multiple CDW orders, and the competition between CDW and SC in detail.
\end{abstract}


{\maketitle}

The recent observation of charge ordering in cuprate high-temperature superconductors\cite{Ghiringhelli2012,Neto2014} has reignited
interests in CDW and its interplay with SC. A new charge ordering is always introduced by different types of instabilities, such as
lattice distortion or FS nesting. However, the driving force behind the CDW phase is still under debate.\cite{Gruner1988,Gruner1994,
Mazin2006,Mazin2008,Weber2011,Yao2006,Mansart2012,Inosov2008,Rossnagel2011,Dai2014,Arguello2015,Laverock2005,Kawasaki2015} From a
Peierls perspective, in an ideal one-dimensional (1D) system, the electronic susceptibility would develop a logarithmic divergence
singularity at some sheets of the FSs spanning by nesting vectors via EPCs, and hence results in a phase transition to the CDW ground
state accompanied by the commensurate/incommensurate periodic lattice distortions and the opening of energy gaps at {\EF}.\cite{
Gruner1988,Gruner1994} Although, the quasi two-dimensional (2D) materials have a weaker tendency towards the nesting-driven CDWs
owing to the imperfect nesting caused by the increased FS curvature, the electronic susceptibility could still be enhanced sufficiently
for a CDW to develop under favorable nesting conditions and EPCs.\cite{Gweon1998,Brouet2004,Brouet2008,Moore2010} Therefore, the CDW
states in quasi 2D systems are particularly attractive due to the existence of possible multiple nesting properties and interacting
collective orders added by the extra dimensionality.

The fairly simple electronic structure of {\RT}, where the CDW instabilities usually develop in the planar square nets of
tellurium,\cite{Laverock2005} provides an unprecedented opportunity to systematically investigate the CDW formation and its
relationship with FS nesting under pretty accurate theoretical models. An incommensurate CDW modulation characterized by a
wave vector $\bm{q}_1 \approx$ 2/7$\bm{c}^*$ was commonly observed.\cite{DiMasi1995,Christos2006,Ru2006,Ru2008} Recently, a
second CDW transition occurs at lower temperatures with $\bm{q}_2 \approx$ 1/3$\bm{a}^*$ perpendicular to the first was discovered
in heavier members of {\RT}.\cite{Ru2008,Moore2010,Hu2011} Additionally, by the Pd intercalation, the suppression of CDWs and even
the emergence of SC were observed.\cite{He2013} Thus, this system offers the valuable possibility to explore the interplay between
multiple CDW instabilities, and also between the CDW and SC orders. However, to date, little is known about the relationship of these
orders belong to different collective phases. In order to obtain a much more comprehensive insight into the multiple CDWs formation
and the interplay between various correlated electronic states, we performed high-resolution ARPES experiments on series of {\PHT}
single crystals with the help of electrical transport measurements, focusing on the evolution of electronic structures and EPCs upon
Pd intercalation. We provide the first systematic electronic structure study on the interplay between multiple CDWs, and their
relationship with SC in {\RT} family as a function of chemical intercalation.

\begin{figure}[htb]
  \begin{center}
    \includegraphics[width=0.75\columnwidth]{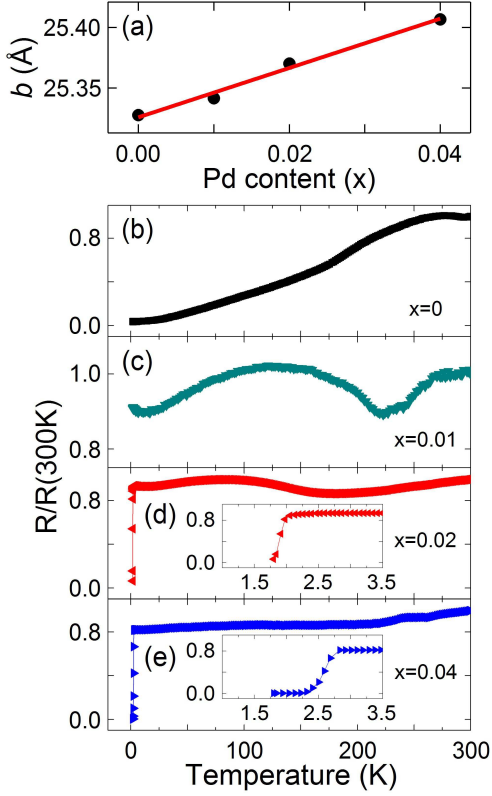}
  \end{center}
  \vspace{-1.3em}
  \caption{(Color online)
  (a) Lattice parameter $b$ as a function of $x$. The straight line is a guide to the eye.
  (b)-(e) Normalized temperature-dependent resistivity data measured on $x$ = 0, {\cdw}, {\cdwsc}, and {\SC}, respectively.
  Insets of (d),(e) are zoom in of $x$ = {\cdwsc} and {\SC} at low temperatures, respectively, showing the superconducting transitions.
  }\label{fig1}
\end{figure}

{\PHT} single crystals with various nominal intercalated compositions ($x$ = 0, {\cdw}, {\cdwsc}, and {\SC}) are used in our study. In
this paper, two incommensurate CDWs are identified. We observe the coexistence of two CDW gaps created by perpendicular nesting vectors
in {\HT} at low temperature. The momentum dependence of the gaps along the inner FS reveals orthorhombicity. A slight Pd intercalation
($x$ = {\cdw}) leads to remarkably different trends of these gaps, the gap symmetry turns out to be near tetragonality, proving the
competition between the CDW orders. Both CDWs are further suppressed in $x$ = {\cdwsc}, leading to the appearance of SC, and they vanish
in $x$ = {\SC}, where the SC order reaches the maximum, showing the competition between CDW and SC. By quantitatively inspecting the
evolution of electronic structures and EPCs, we demonstrate the nesting nature of multiple CDWs, suggest the significant increment of
the second CDW gap from $x$ = 0 to {\cdw} determined by the EPC strength. The competition between these two CDWs, and their relationship
with SC is very likely to be a FS competition scenario.

High-quality single crystals of {\PHT} were synthesized by the flux method.\cite{Pfuner2010} ARPES measurements were performed
at Renmin University of China and Institute of Physics, Chinese Academy of Sciences, with a He-discharge lamp, at the 1-cubed
ARPES end station at BESSY and PGM beam line of the Synchrotron Radiation Center (Stoughton, WI). Spectra were recorded with 55
eV photons, taken at $T$ = 30 K, with a pressure better than 4 $\times 10^{-11}$ Torr. Electrical transport measurements were
performed in a PPMS-14 (Quantum Design).

\begin{figure}[htb]
  \begin{center}
    \includegraphics[width=0.85\columnwidth]{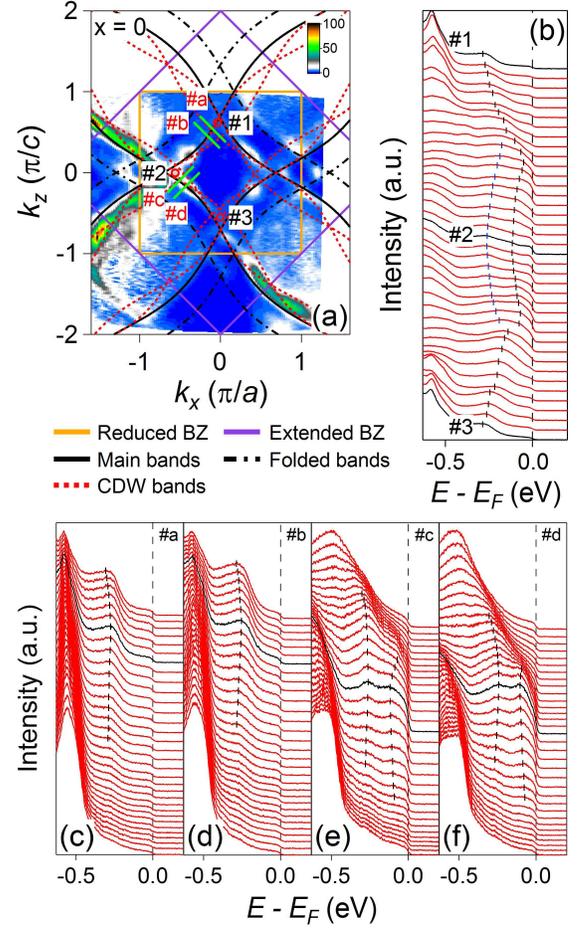}
  \end{center}
  \vspace{-1.3em}
  \caption{(Color online)
  (a) ARPES intensity plot of {\HT} at {\EF} as a function of the 2D wave vector. The intensity is obtained by integrating
      the spectra within $\pm$15 meV with respect to {\EF}. The hopping parameters, $t_\parallel$ and $t_\perp$, of the
      superimposed TB bands are $\sim$1.85 and $\sim$0.50 eV, respectively.
  (b) EDCs measured at various $k_F$ points of the inner FS along \#1$-$\#2$-$\#3, where \#1, \#2, and \#3 are indicated
      by red circles in (a). The spectra at these three $k_F$ points are highlighted by black curves.
  (c)-(f) EDC plots of cuts \#a$-$\#d indicated by green lines in (a), respectively. The CDW gap definitions and corresponding $k_F$
          positions are emphasized by black curves. Black and blue dashes in (b)-(f) are extracted peak positions, serving as guides
          to the eye.
  }\label{fig2}
\end{figure}

\begin{figure*}[htb]
  \begin{center}
    \includegraphics[width=1.83\columnwidth]{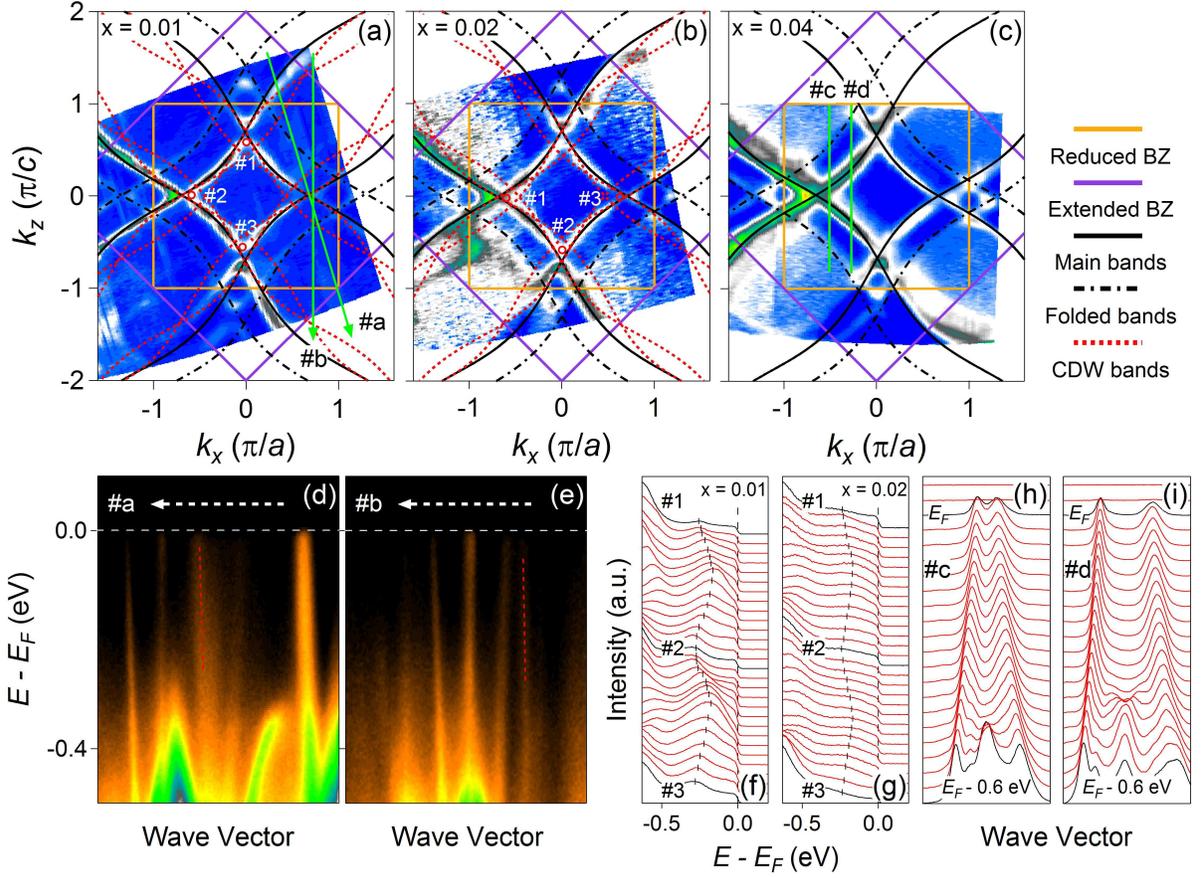}
  \end{center}
  \vspace{-1.4em}
  \caption{(Color online)
  (a)-(c) FSs for $x$ = {\cdw}, {\cdwsc}, and {\SC}, respectively, obtained by integrating the spectra within $\pm$15 meV with
          respect to {\EF}. The superimposed TB bands are calculated with the hopping parameters equally to that of {\HT} and
          fine adjustment of {\EF} values.
  (d),(e) ARPES intensity plots of $x$ = {\cdw} along cuts \#a and \#b, respectively, indicated by green arrows in (a). Red dashes
          are guides to the eye for the shadow bands.
  (f),(g) EDCs measured at various $k_F$ points of the inner FS along \#1$-$\#2$-$\#3, where \#1, \#2, and \#3 are indicated by red
          circles in (a),(b), respectively. The spectra at these three $k_F$ points are highlighted by black curves. Black dashes
          are extracted peak positions, serving as guides to the eye.
  (h),(i) Momentum distribution curve (MDC) plots of $x$ = {\SC} along cuts \#c and \#d, respectively, indicated by green lines in (c).
  }\label{fig3}
\end{figure*}

As presented in Fig.~\ref{fig1}(a), the monotonic increase of lattice constant $b$ along with the increasing Pd content demonstrates
the successful intercalation of Pd into the weakly bonded double Te layers. Figs.~\ref{fig1}(b)-\ref{fig1}(e) show the temperature
dependence of electrical resistivity for {\PHT}. Consistent with previous work,\cite{Ru2008} two bumps can be seen at $\sim$290 and
$\sim$100 K in {\HT}, suggesting two well-separated CDW transitions. The second CDW transition at $\sim$100 K is barely visible in
the resistivity, this was interpreted as a renormalization of the electron dispersion in the ungapped FS parts, which may partially
compensate the opening of the CDW gap on some FS sheets.\cite{Sinchenko2014} With the increasing amounts of Pd ($x$ = {\cdw}), these
two CDWs exhibit opposite trends, two bumps merge at $\sim$220 K. More clearly observed in the insets of Figs.~\ref{fig1}(d) and
\ref{fig1}(e), the SC emerges in $x$ = {\cdwsc} around 2.0 K with further suppressed CDW orders, and reaches the maximum ($\sim$2.8
K) in $x$ = {\SC} accompanied by the vanished CDW instabilities, respectively.

We present in Fig. 2 the electronic structure of {\HT}. The FS in Fig.~\ref{fig2}(a) is well described by a 2D tight-binding (TB)
model including only the in-plane $p_x$ and $p_z$ orbitals of a Te plane, except at the crossings between them, owing to their
interactions are neglected in our calculation. The dispersions for $p_x$ and $p_z$ can be readily derived as,
 $$ E_{p_x}(\textbf{k}) = -2t_{\parallel}\cos [(k_x + k_z)\frac{a}{2}] + 2t_{\perp}\cos [(k_x - k_z)\frac{a}{2}] - E_F, $$
 $$ E_{p_z}(\textbf{k}) = 2t_{\perp}\cos [(k_x + k_z)\frac{a}{2}] - 2t_{\parallel}\cos [(k_x - k_z)\frac{a}{2}] - E_F. $$
Details of the TB model are described elsewhere.\cite{Yao2006} The residual spectral distribution along the inner pocket, which
is similar to that in ErTe$_3$,\cite{Moore2010} indicates the existence of multiple CDWs characterized by different nesting vectors.

\begin{figure}[htb]
  \begin{center}
    \includegraphics[width=0.98\columnwidth]{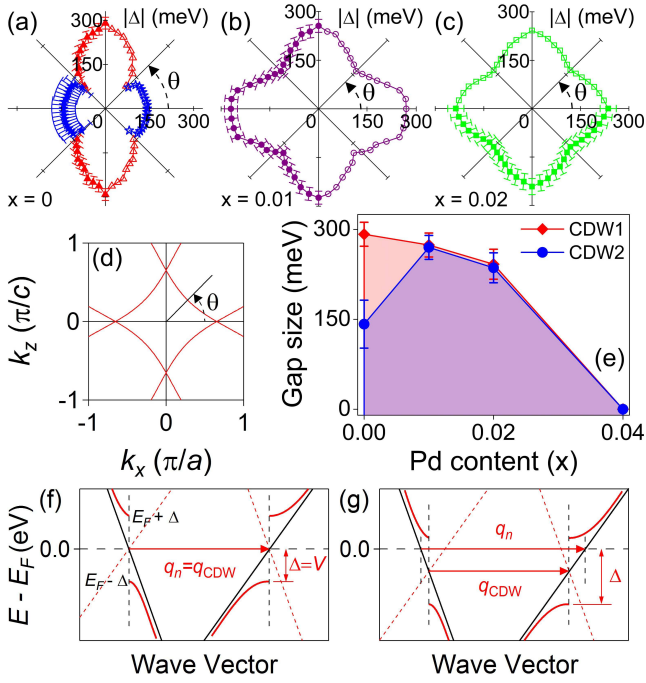}
  \end{center}
  \vspace{-1.35em}
  \caption{(Color online)
  (a)-(c) Polar plots of the CDW gap size for the inner FS of $x$ = 0, {\cdw}, and {\cdwsc}, respectively, as a function of
          FS angle ($\theta$) defined in (d). Filled symbols with error bars are the original data, consecutively extracted
          from the EDCs in Figs.~\ref{fig2}(b), \ref{fig3}(f), and \ref{fig3}(g). Open symbols are the folded data which take
          into account the orthorhombic symmetry.
  (d) Schematic inner FS and definition of the FS angle ($\theta$).
  (e) Summary of the two CDW gap magnitudes as a function of Pd intercalation, the values are extracted at the corners of the
      inner FSs.
  (f),(g) Schematic pictures illustrating two-band CDW intersection for {\PHT} when $\bm{q}_{n}$ = $\bm{q}_{\mathrm{CDW}}$ and
          $\bm{q}_{n}$ $\textgreater$ $\bm{q}_{\mathrm{CDW}}$, respectively. $\Delta$ mixes two states in the unperturbed bands
          (thick black lines) connected by $\bm{q}_{\mathrm{CDW}}$, resulting in the gapped CDW bands indicated by thick red
          curves. The red dashed lines are translated bands by $\pm\bm{q}_{\mathrm{CDW}}$.
  }\label{fig4}
\end{figure}

We choose two sets of cuts (\#a,\#b and \#c,\#d) to elucidate the different appearance of multiple CDW gaps, and present
the corresponding energy distribution curves (EDCs) in Figs.~\ref{fig2}(c)-\ref{fig2}(f). Both cuts \#a and \#b show a
back-bending band feature, this is reminiscent of the dispersion of Bogoliubov quasiparticles, suggesting the opening of
a CDW gap. Instead, cuts \#c and \#d reveal both two quasiparticle branches, demonstrating the center of the CDW gap is
pushed below {\EF} due to the longer than perfect nesting vector observed.\cite{Gruner1994,Moore2010} As illustrated in
Figs.~\ref{fig4}(f) and \ref{fig4}(g) for {\PHT}, the definition of CDW gap is dependent on both the nesting conditions
and EPC strength. For an ideal nesting, the CDW gap center locates at {\EF}, thus only the lower branches are visible;
for an imperfect nesting with $\bm{q}_{n}$ $\textgreater$ $\bm{q}_{\mathrm{CDW}}$, the gap center is pushed below {\EF},
moreover, if the coupling parameter ($V$) between these two nested states including EPCs is smaller than the binding
energy of gap center, both two quasiparticle branches can be observed.\cite{Gruner1988,Mazin2008,Brouet2008,Yao2006,
Voit2000} Reasonably, the CDW gap definitions can be classified as, the gap between the lower branch and {\EF} is used
for the situations with only lower branches visible, and the gap between these two branches is for the latter case,
yielding 2$V$. More details are discussed quantitatively below.

This apparent distinction indicates a second CDW gap different from that in cuts \#a and \#b. Detailed inspection on the
binding energies of the branches at $k_F$ reveals the decreasing trend from \#a ($\sim$0.275 eV) to \#b ($\sim$0.255 eV),
and the same trend holds from \#c ($\sim$0.121 eV) to \#d ($\sim$0.103 eV), manifesting the shortened nesting vectors.
To precisely determine the gap size and their momentum dependence, the EDCs at various $k_F$ points of the inner FS along
\#1$-$\#2$-$\#3 are displayed in Fig.~\ref{fig2}(b). Corresponding to the first gap definition, one can obtain a large CDW
gap near the corners \#1 and \#3 with only lower branches observed, which gradually decreases off the corners due to the
imperfect nestings. And according to the second definition, a small CDW gap appears around \#2 with its center below {\EF},
which approaches {\EF} when off the corner, and both two quasiparticle branches visible. These behaviors are consistent
with the nesting-driven scenario.\cite{Gruner1994,Laverock2005,Yao2006,Ru2008}

To illustrate the characteristic of multiple CDWs and fully understand the interplay between them, we elaborate on the electronic
structure evolution upon Pd intercalation. Figs.~\ref{fig3}(a)-\ref{fig3}(c) show the FS mapping data of $x$ = {\cdw}, {\cdwsc},
and {\SC}, respectively. One can see the nearly fourfold symmetric intensity variation in the FSs of $x$ = {\cdw} and {\cdwsc}
rather than the apparent twofold symmetry in {\HT}, manifesting the symmetry of gap-opening connected to a CDW picture.\cite{
Brouet2004,ResidualFS} The shadow bands sketched by red dashes in Figs.~\ref{fig3}(d) and \ref{fig3}(e), indicated via cuts \#a
and \#b in Fig.~\ref{fig3}(a), respectively, also confirm the validity of FS nesting in determining the CDW instabilities in
this system.

The detailed gap anisotropy study of $x$ = {\cdw} and {\cdwsc}, like that of {\HT} in Fig.~\ref{fig2}(b), are displayed in
Figs.~\ref{fig3}(f) and \ref{fig3}(g), respectively, revealing slightly orthorhombic, or nearly tetragonal symmetry, in stark
contrast to the orthorhombicity in {\HT}. In the nesting-driven CDW picture, the origin of the intensity variation along the
FS can be reasonably interpreted by the momentum dependence of these CDW gaps. We summarize the gap size along the inner FS
of $x$ = 0, {\cdw}, and {\cdwsc} as a function of the FS angle ($\theta$) in Figs.~\ref{fig4}(a)-\ref{fig4}(c), respectively,
clearly confirming the anisotropy. It is noted that, as indicated by blue filled pentacles in Fig.~\ref{fig4}(a), the definition
of the second CDW gap in {\HT} is different from others' ascribed to the observation of both upper and lower quasiparticle branches.
According to the different gap definitions discussed above, instead of the gap between the lower branch and {\EF} for others, we
use the one between these two branches, characterized by 2$V$.\cite{Gruner1988,Mazin2008,Brouet2008,Yao2006,Voit2000}

\begin{figure}[htb]
  \begin{center}
    \includegraphics[width=0.83\columnwidth]{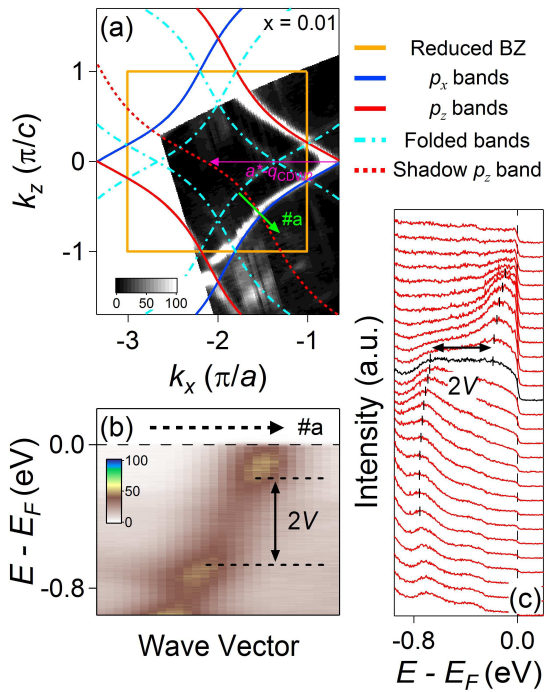}
  \end{center}
  \vspace{-1.35em}
  \caption{(Color online)
  (a) FS for $x$ = {\cdw} around the second 3D (reduced) BZ center obtained by integrating the spectra within $\pm$15 meV
      with respect to {\EF}. The calculated TB bands with orbital projection and shadow $p_z$ band are indicated to enable
      a study of the coupling strength between the states linked by a nesting vector, $\bm{a}^* - \bm{q}_{\mathrm{CDW2}}$,
      shown as magenta arrow.
  (b) ARPES intensity plot measured at the crossing between $p_x$ and shadow $p_z$ bands along cut \#a, indicated by green
      arrow in (a), and (c) corresponding EDC plot. The definition of 2$V$ is emphasized by black curve. Black dashes are
      extracted peak positions, serving as guides to the eye.
  }\label{fig5}
\end{figure}

We now discuss the nesting properties associated with the multiple gaps and spectral distribution along the FS. By comparing the
nesting vectors determined by ARPES and TB model calculations with the x-ray diffraction results,\cite{Ru2008} we can reasonably
interpret the momentum dependence of the multiple CDW gaps and the gap center of the second CDW in {\HT} underlying {\EF}. For {\HT},
the two nesting vectors reveal $\bm{q}_{n1}$ = $\bm{c}^* - \bm{q}_{\mathrm{CDW1}} \approx$ 0.71(3)$\bm{c}^*$ and $\bm{q}_{n2}$ =
$\bm{a}^* - \bm{q}_{\mathrm{CDW2}} \approx$ 0.69(5)$\bm{a}^*$, where the former is in complete agreement with the perfect nesting
vector along $c$ axis and the latter is longer than that along $a$ axis, respectively. Thus, the center of the second CDW gap is
pushed below {\EF} even at the corner (\#2) of the inner FS, and gradually moves to lower binding energy with the shortened nesting vector.\cite{Ru2008,Gruner1994,Brouet2008} For $x$ = 0.01 and 0.02, the ones parallel to $c$ axis are 0.68(8) and 0.67(1)$\bm{c}^*$,
and to $a$ axis are 0.68(4) and 0.67(3)$\bm{a}^*$, respectively, elucidating the FS topology becomes more isotropic upon Pd intercalation.

The evolution of the maximal gap size of the two CDW orders upon Pd intercalation are summarized in Fig.~\ref{fig4}(e).
Combining the multiple nesting properties discussed above, it is conspicuous that upon Pd intercalation, the interplay
between these two collective phases yields competition, similar to the optical spectroscopy results upon chemical
pressure,\cite{Hu2014} and to the transport measurements under pressure.\cite{Hamlin2009,Zocco2015} These can to some
extent be explicated by our multiple nesting properties, but not adequate yet. Along with the increasing Pd content,
the first CDW gap gradually decreases ascribed to the shortened nesting vector. Simultaneously, the second CDW gap
substantially increases for $x$ = {\cdw}, and then decreases monotonically. The appreciable transform of the second
CDW gap magnitude from $x$ = 0 to {\cdw} strongly suggests that, upon Pd intercalation, not only the FS evolution
pointing to the nesting picture, but also the variation of EPC strength is needed to be fully included for investigating
the interplay between multiple CDW orders quantitatively.

According to the above discussions, one can obtain the coupling strength ($V$) between the states linked by $\bm{a}^* -
\bm{q}_{\mathrm{CDW2}}$ in {\HT} from Fig.~\ref{fig2}(b), where 2$V$ = 0.142 eV. However, as only the lower quasiparticle
branch being observed in $x$ = {\cdw}, one cannot determine the $V$ unambiguously. Thus, we perform the measurements extended
to the second three-dimensional (3D) BZ on $x$ = {\cdw}, and illustrate the FS in Fig.~\ref{fig5}(a). The dispersion presented
in Fig.~\ref{fig5}(b), indicated via cut \#a in Fig.~\ref{fig5}(a), is measured at the crossing between $p_x$ and shadow $p_z$
bands, of which the intensity is too weak to be clearly visible in the plot. It is distinct that a gap completely opens below
{\EF}. Since these two states are coupled by a nesting vector, guided by magenta arrow in Fig.~\ref{fig5}(a), the interaction
between them would also yields the $V$.\cite{Brouet2008} We now can precisely measure 2$V$ using the corresponding EDC plot
in Fig.~\ref{fig5}(c), showing 2$V$ = 0.484 eV. Based on this, it is further confirmed that the EPC strength undergoes a
significant variation from $x$ = 0 to {\cdw}.

The detailed studies on the evolution of electronic structures and EPCs give us a high chance to extract the nature of the
interplay between multiple CDWs, as well as their relationship with the SC in {\PHT}. According to the temperature versus
pressure phase diagram proposed in Refs. 31 and 32, the role of Pd intercalation on the evolution of two CDWs and SC is
similar to that of pressure performed on {\RT} compounds. As already pointed out, the first CDW is suppressed monotonically
because of the imperfect nesting. Simultaneously, the EPC strength of the second CDW has a remarkable transform from $x$ =
0 to {\cdw}. This may be driven by the possible lattice distortion towards the tetragonal structure upon Pd intercalation,
resulting from the competition between CDW orders for the low-energy spectral weight available for nesting.\cite{Hu2014}
Nevertheless, due to the longer than perfect nesting vector along $a$ axis, the center of the second CDW gap still slightly
locates below {\EF} in $x$ = {\cdw}. Thus, the gap further decreases even with an ideal nesting vector in $x$ = {\cdwsc}. The
competition and suppression of these two CDW instabilities give rise to the emergence of SC in $x$ = {\cdwsc}, then the SC order
reaches the maximum with the vanished CDWs in $x$ = {\SC}. Extensive work has been carried on the interplay between CDW and SC,
principally in cuprate high-temperature superconductors,\cite{Ghiringhelli2012,Neto2014} yet the underlying mechanism for the
competition is still controversial. Our ARPES results for the relationship between these collective states may shed light on
it, even the high-temperature SC. As shown in the phase diagram of polycrystalline {\PHT} in Ref. 25, the weak Pd-intercalation
dependence of $T_c$ indicates the SC possibly not determined by quantum critical fluctuations, the competition for FS with CDWs
based on the Bilbro-McMillan partial gaping scenario may be the dominant nature instead.\cite{Hamlin2009,Bilbro1976}

To conclude, we have performed ARPES and electrical transport experiments on {\PHT} single crystals to study the interplay
between multiple CDWs and their relationship with SC. We report the systematic evolution of the electronic structures and
EPCs upon Pd intercalation, determine the nesting-driven nature of the CDWs formation, and find the competition between
these CDW orders is for the low-energy spectral weight. The compelling evidences for the dramatic transform of EPC strength
along $a$ axis from $x$ = 0 to {\cdw} leave the effect of Pd intercalation further complicated, requiring future studies to
clarify. The competition between SC and CDWs for the FS may provide insight into the microscopic origin of high-temperature
SC, paving the way to identify more high-temperature superconductors.

We would like to thank Anmin Zhang, Yong Tian, and Qingming Zhang for helpful discussions, and Hechang Lei for the help in
plotting figures. This work was funded by grants from the National Natural Science Foundation of China (Nos. 11274381 and
11404175), the National Basic Research Program of China (973 Program), the Ministry of Education of China, and the Chinese
Academy of Sciences (No. XDB07000000).

\end{document}